\documentclass[preprint,showpacs,preprintnumbers,amsmath,amssymb,floatfix]{revtex4-1}

\usepackage{color}
\usepackage{graphicx}
\usepackage{dcolumn}
\usepackage{bm}
\usepackage{multirow}
\definecolor{sienna}{cmyk}{0,0.72,1,0.45}
\definecolor{fg}{cmyk}{0.91,0,0.88,.12}
\definecolor{yellow}{cmyk}{0,0,1,0}
\definecolor{or}{cmyk}{0,1,0.5,0}
\definecolor{magenta}{cmyk}{0,1,0,0}
\definecolor{rubinered}{cmyk}{0,1,0.13,0.45}
\definecolor{blue}{cmyk}{1,1,0,0}
\definecolor{turquoise}{cmyk}{1,1,0,0.5}
\definecolor{aquamarine}{cmyk}{0,1,0,0.0}
\definecolor{midnightblue}{cmyk}{1,0.5,0.0,0.0}
\definecolor{junglegreen}{cmyk}{1,0,0.2,0.5}

\begin{document}

\title{Transmission of packets on a hierarchical network: Statistics and explosive percolation}

\author{Ajay Deep Kachhvah}
\email{ajay@physics.iitm.ac.in}
\affiliation{Department of Physics, Indian Institute of Technology Madras, India.}
\author{Neelima Gupte}
\email{gupte@physics.iitm.ac.in}
\affiliation{Department of Physics, Indian Institute of Technology Madras, India.}

\begin{abstract}
We analyze an idealized model for the transmission or flow of particles,
or discrete packets of information, in a weight bearing branching hierarchical $2-D$ networks, and its variants. The capacities add hierarchically down the clusters. Each node can accommodate a limited number of packets, depending on its capacity and the packets hop from node to node, following the links between the nodes. The statistical properties of this system are given by the Maxwell - Boltzmann distribution. We obtain analytical expressions for the mean occupation numbers as functions of capacity, for different network topologies. The analytical results are shown to be in agreement with the numerical simulations. The traffic flow in these models can be represented by  the site percolation problem. It is seen that the percolation transitions in the $2-D$ model and in its variant lattices are continuous transitions, whereas the transition is found to be explosive (discontinuous) for the $V-$ lattice, the critical case of the $2-D$ lattice. The scaling behavior of the second order percolation case is studied in detail. We discuss the implications of our analysis.


\end{abstract}

\maketitle

\section{Introduction}

Studies of dynamical behavior on complex networks have witnessed
exponential growth in recent years. This rapid growth has been fueled
by the application of network science to diverse areas like biology,
sociology, physics and computer science and others
 \cite{strogatz,barabasi,newman}. The study of the flow of traffic
through a network is a important and frequently encountered dynamical
process. The dynamics of the flow of information has been studied over a
wide range of communication networks of various types
\cite{csabai,takayasu,radner,ohira,trety,guimera,brajendra,mukherjee},
with the Internet being an important example. In
the internet, computers of different computational processing capacities
communicate by exchanging information in the form of discrete units,
or ``packets'' which then  find their destination by hopping from node to node via the links between them. Thus, the transmission of information on the network is mapped onto the traffic of packets or
particles moving in the network via connections. The packets or
particles are often assumed to be non-interactive, however, to
understand phenomena like congestion, the interaction between the
packets has to to be taken into account. In real networks, the finite
processing capacity or processing power of nodes causes collective
behavior such as congestion, jamming or the failure of transmission of
packets. The effect of congestion in networks has been
studied via centrality measures such as the betweenness centrality
\cite{kahng,goh,goh2,barth,ghim,newman2}, as well as via order
parameters like the fraction of packets delivered \cite{tadic,manna,brajendra,mukherjee}. Cellular automata models have also been useful in studying the dynamics of the transmission of particles or packets in the network \cite{barth,ghim,newman2,barth2,sole,monero,holme,tadic, steger,brajendra,solla,zhao}.

An interesting direction of analysis for packet transport has been to explore the stationary
distributions of particles at various nodes, and to map them on to the
usual distributions of statistical ensembles \cite{moura,germano}. This
has been carried out both for lattices of homogeneous capacity at
nodes and for the more realistic case of inhomogeneous capacity at nodes.
The model \cite{moura}, in which the particles move randomly in the
network with the constraint that each node can be occupied by at most
one particle, conforms to the Fermi -Dirac distribution. Another model \cite{germano}, in which
particles performing a random walk on nodes of finite processing capacity with
the constraint that no particle can move to a node which has already has
its maximum permitted number of packets, satisfies the
Maxwell-Boltzmann distribution for the statistics of particles.
Most of the earlier studies of the transport of packets have been carried out for
scale-free or random networks \cite{csabai,ohira,trety,guimera,brajendra,mukherjee}, apart from some studies of transport in small world networks \cite{latora}. However, very little analysis has been carried out for the transport of packets in tree- or river- like branching hierarchical networks \cite{basumohanty}. In this paper, we explore the transport dynamics and statistics on a hierarchical regular $2D$ lattice model, two of its variants \cite{janaki}, and its critical case \cite{ajay}. The base $2D$ model is, in fact, the $q(0,1)$ case of the Coppersmith model \cite{copper}. Analogous models are river network models \cite{river}, voter models \cite{griffeath}, the Domany Kinzel cellular-automata model \cite{domany} and the branching hierarchical model of the lung inflation process \cite{suki,suki2}. In all the four lattice networks analyzed by us, the nodes bear finite processing capacity and the packets hop from node to node utilizing the lattice links, incorporating the constraint that no packet can hop to a node that has already has its maximum allowed number of packets. If the capacity distribution of a node is mapped to energy levels, the packet dynamics of transport in these hierarchical networks is found to follow the Maxwell - Boltzmann distribution. 

The transport of packets on a network is also analogous to the site
percolation problem. When the percolating network supports a giant cluster, then communication is possible through this large connected component, otherwise communication breaks down. Recently, Achlioptas $et$
$al.$ \cite{achliop} studied percolation for the Erdos Renyi model using
a product rule and found that the giant component emerged suddenly at
the percolation threshold, and that the percolation transition was discontinuous. This discontinuous percolation transition appears when the growth of the largest cluster is systematically suppressed
thereby promoting the formation of several large components that eventually merge in an explosive way \cite{friedman}. Several aggregation models, based on percolation, have been developed to achieve
this change in the nature of transition \cite{achliop,radicchi,ziff,araujo,chen,moreira,chao,dsouza,manna}. Recent numerical studies argue that the transition is really a continuous transition, with the discontinuity seen being due to the finite size of the lattice \cite{daCosta}. For the Achlioptas case, rigorous mathematical arguments confirm that the transition in the model, is really continuous \cite{riordan}. However, other models show what appears to be a genuine first order transition \cite{araujo, chen}. 

The explosive percolation transition is seen in our hierarchical models in an unusual context.
The percolation transitions in our base $2D$ model, and its two variants, have also
been explored and  are found to be the usual continuous transition. However, the percolation
transition in the $V-$ lattice, the critical case of the $2D$ model is found to be discontinuous one. In this paper we provide numerical evidence for the discontinuity of percolation transition observed in the $V-$ lattice. This explosive transition does not follow the scaling laws of the continuous percolation transition. It is interesting to note that the explosive percolation transition is seen in only one realization, viz. the $V-$ lattice or the critical realization of the base lattice. This realization is also the only case where power-law scaling is observed in the avalanche distributions \cite{aval}. Thus the  structure of the $V-$ lattice appears to contribute to the critical behavior of transport processes on the structure. The analysis of this case may lead to general insights into the behavior of the explosive percolation transition, which has not been explored for
hierarchical networks.

\section {The branching hierarchical network and its variants}

\begin{figure}[htb]
\begin{center}
\begin{tabular}{cc}
\includegraphics[height=4.5cm,width=6.5cm]{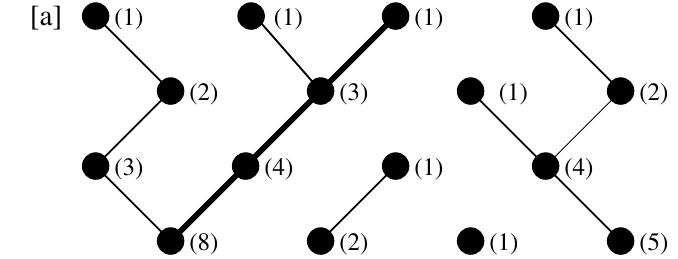}&
\hspace{0.5cm}
\vspace{0.75cm}
\includegraphics[height=4.5cm,width=6.5cm]{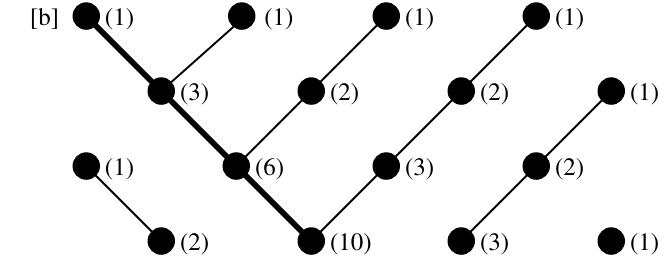}\\
\includegraphics[height=4.5cm,width=6.5cm]{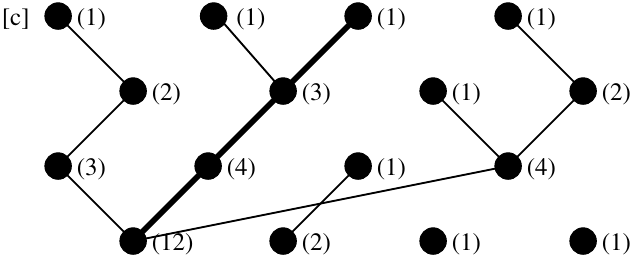}&
\hspace{0.5cm}
\vspace{0.75cm}
\includegraphics[height=4.5cm,width=6.5cm]{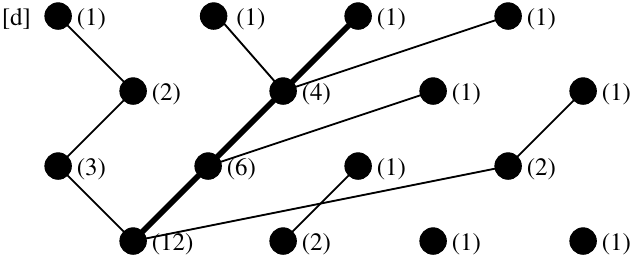}\\
\end{tabular}{}
\caption{\label{fig:variants} (a) The original lattice, (b) the $V-$ lattice, (c) the strategy-I lattice and (d) the strategy-II lattice of $L=M\times M$ sites, where $M=4$. The numbers in brackets beside each site denote the weight bearing capacity of that site. The thick line in each panel denotes the trunk of the respective lattice (network).}
\end{center}
\end{figure}
We first discuss the branching, hierarchical networks studied in this paper.
The original network was set up as a load bearing network,
and connectivity strategies were devised to enhance its load bearing properties \cite{janaki}.
Here, we describe the base $2D$ lattice network model, a critical case of the base lattice, the $V-$
lattice \cite{ajay}, and its two variant load enhancing strategies, strategy-I and strategy-II
\cite{janaki}. These are shown in the four panels (a), (b), (c), and (d) of Fig. \ref{fig:variants} respectively. The base lattice network model has similarities with diverse models of real-world networks like the river network \cite{river} model, the voter model \cite{griffeath}, the cellular-automata model \cite{domany}, and the lung inflation model \cite{suki, suki2}.

The  lattice shown in panel \ref{fig:variants}(a), which will be referred to as the original lattice henceforth,  is, in fact, a critical case of the Coppersmith model for granular media \cite{copper}, and also a model for the river network \cite{river}. The lattice is a regular triangular lattice of sites with each site in the top layer having unit capacity. A site $i$ at some layer $D$ is connected to one of its two nearest neighbor sites chosen at random in the layer $D+1$. The total capacity of site $i$ in the layer $D$ is transmitted to its left neighbor $i_l$, or right neighbor $i_r$, in the layer $D+1$. Thus, the capacity of  the site in the layer $D$ at the $i$th site, $w(i^D)$, satisfies the stochastic equation
\begin{equation}
w(i^D)=l(i^{D-1}_l,i^D)w(i^{D-1}_l)+l(i^{D-1}_r,i^D)w(i^{D-1}_r)+1,
\end{equation}
where $D=1,2,\ldots,M$ with $M$ being the lattice side or the total number of layers in the network. The quantity $l(i^{D-1}_l,i^D)$ in Eq. $1$ takes the value $1$ if a connection exists between ${\it i^{D-1}_l}$ and ${\it i^{D}}$, and $0$ if otherwise. The network consists of many clusters, where a cluster is defined as a set of sites connected with each other. The largest cluster in the system is called the maximal cluster, and its trunk or backbone is the set of connected sites with the highest weight-bearing capacity in the maximal cluster.

The original lattice has a very special realization which has the
maximum trunk strength or capacity compared to all other possible
realizations. This lattice bears a unique $V-$ shaped cluster that
includes all the sites in the first layer, and $(M-I+1)$ sites in the
$I-$th layer. One of the arms of the $V$ constitute the trunk, and all other connections run parallel
to the arm of the $V$ that is opposite to the trunk. Thus, this cluster
includes the largest number of sites, and is thus the largest possible
cluster the original lattice could have. We call this lattice the $V-$
lattice (panel \ref{fig:variants}(b)), and the cluster, the $V-$ cluster. 
Structures similar to the $V-$ lattice can be seen in riverine deltas \cite{riverdelta}, in Martian gullies \cite{marsgully}, and in granular flows \cite{shinbrot}, if the channels of maximal flow capacity are considered \cite{assym}.   
The other two variants of the original lattice which we have studied for
simulation are the strategy-I (panel \ref{fig:variants}(c)) and the strategy-II (panel \ref{fig:variants}(d)) lattices. These two strategies \cite{janaki} were developed for strengthening the load bearing capacity of the
lattice by reconnecting a site on a given layer to a suitable site in
the layer below with the reconnections restricted to at most one per
layer. In strategy-I, the maximum number of sites are included in the
maximal cluster by connecting as many disjoint clusters as possible to
the sites on the trunk $T$ of the maximal cluster. This is a bottom to
top strategy. Another strategy is a top to bottom strategy where the reconnections start from the first layer onwards is also possible, so that the capacities of the site on the trunk in the layer below and its subsequent sites get enhanced in every layer. This is strategy -II (see \cite{janaki} for further details).
\begin{figure}[htb]
\begin{center}
\begin{tabular}{cc}
\includegraphics[height=5cm,width=6.5cm]{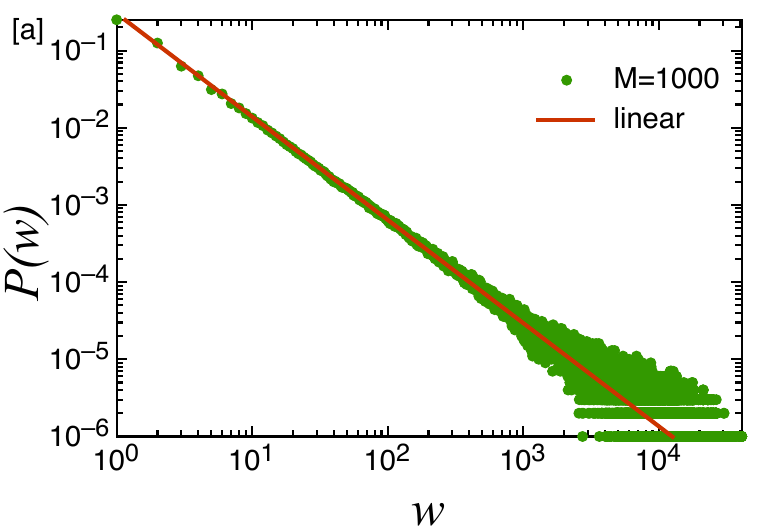}&
\includegraphics[height=5cm,width=6.5cm]{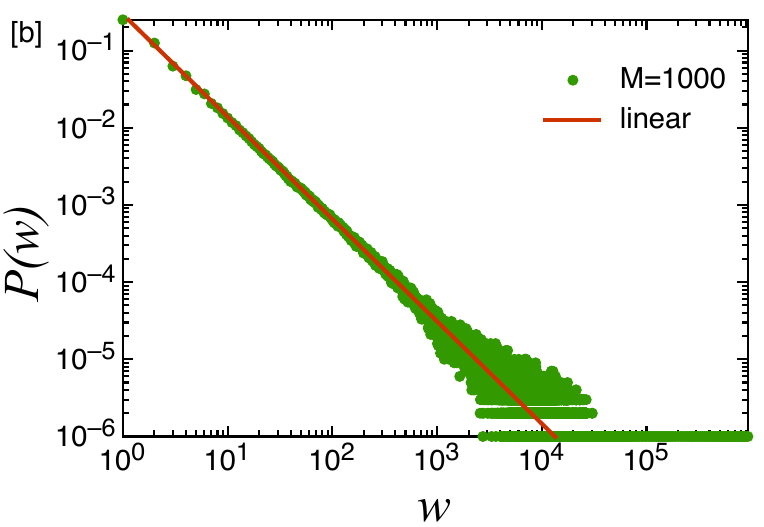}\\
\includegraphics[height=5cm,width=6.5cm]{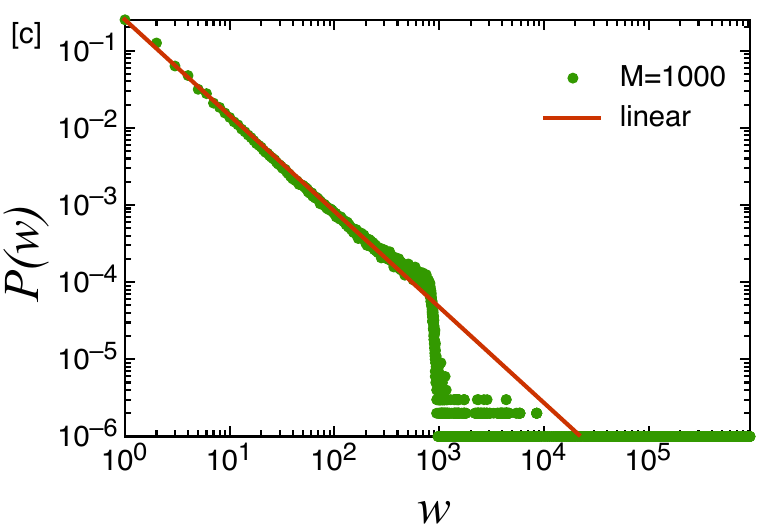}&
\includegraphics[height=5cm,width=6.5cm]{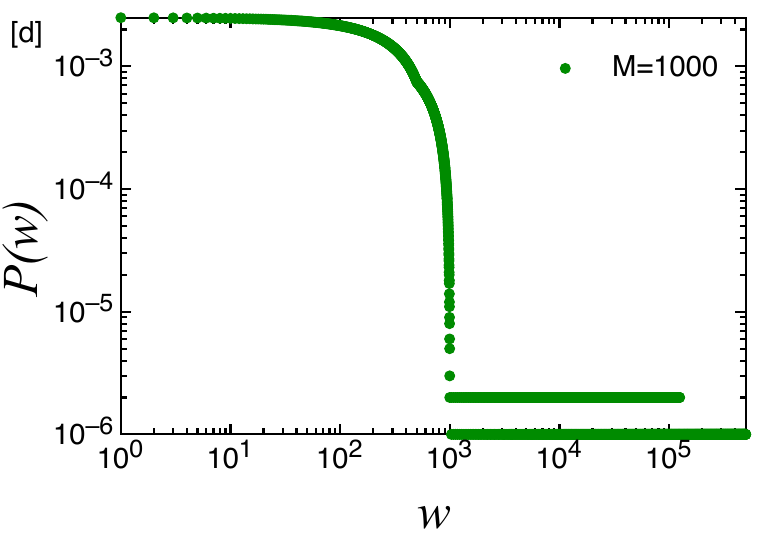}\\
\end{tabular}{}
\caption{\label{fig:capdist} (Color online) The distributions of capacities $w$ displaying power law behavior ${P(w)=cw^{-\alpha}}$ for (a) the original lattice with exponent $\alpha=1.3335$ and $\chi^2=9.3924$, (b) the strategy-I with exponent $\alpha=1.336$ and $\chi^2=3.3028$, and (c) the strategy-II with exponent $\alpha=1.2309$ and $\chi^2=0.0924$. The capacity distribution for (d) the $V-$ lattice does not display power law behavior. Here, lattice size is $L=M\times M$, where $M=1000$.}
\end{center}
\end{figure}

We have carried out simulations for the capacity distribution $P(w)$ of all
the four lattice networks mentioned above. Here the capacity distribution
is, in fact, the probability that a site has capacity $w$ over
the entire network. The distributions of capacities for the original,
the strategy-I and the strategy-II network are governed by a power
law of the form ${P(w)=cw^{-\alpha}}$, where $c$ is a proportionality constant. However, the capacity distribution for the $V$ - lattice does not behave in accordance with the power law. The exponents of the
power law for the original, the strategy-I, the strategy-II lattices are
found to be $1.333$, $1.336$, and $1.23$ respectively as shown in the
panels \ref{fig:capdist}(a), \ref{fig:capdist}(b), and
\ref{fig:capdist}(c) of Fig. \ref{fig:capdist}. The mean-field analysis for the capacity
distributions for these three networks has been carried out by Kachhvah $et$ $al.$
\cite{ajay} using the method of Takayasu \cite{takayasu} for the
original lattice.

\section {Statistics of packet transmission on the networks}

We now consider packet transport on the hierarchical lattices.
A packet or particle following the topology of the network hops from one site to a neighboring site along the link between them. In this scenario, the time the packet spends in the given site is directly proportional to the capacity of that
site $i.$$e.$ the probability $p_i$ of a packet to be found in a site $i$ of branching hierarchical network, is proportional to the capacity $w_i$ of that site, $i.$ $e.$
\begin{equation}
p_i=\frac{w_i}{\sum_i w_i},
\end{equation}
where the sum is over all the nodes of the hierarchical network.

In statistical physics, we know that the occupation probability of a site $i$ with energy $\varepsilon_i$ is proportional to the Boltzmann factor
\begin{equation}
p_i=C^{\prime} \mathrm{e}^{-\beta \varepsilon_i},
\end{equation}
where $C^{\prime}$ is a normalization constant, $\beta$ plays the role of the inverse temperature, $\beta=1/T$, and $\varepsilon_i$ is the energy associated with the site $i$ with capacity $w_i$ in the network.
Equating the $p_i$ in Eqs. $(1)$ and $(2)$, the relation between the energy associated with a site $i$, and its capacity $w_i$ is given by \cite{foot}
\begin{eqnarray}
\mathrm{e}^{-\beta\varepsilon_i}=\it{Cw_i},
\end{eqnarray}
where $C$ is a constant.
If $n_i$ be the mean occupation number of node $i$ with its capacity
$w_i$ and where $n_i\leq w_i$, the total number of packets or particles
$N$ \cite{huang} present on the network would then be 
\begin{equation}
N=\sum_{i} \big<n_{i}\big>.
\end{equation}
The total energy $E$ of the network is then given by $E=\sum_i \varepsilon_i n_i$.
Since the packets hopping on the sites in the hierarchical network are distinguishable from each other, 
the canonical partition function $Z_N$ \cite{huang} when $N$ distinguishable packets are present in the system  is then given by
\begin{eqnarray}
Z_N={\sum_{\{n_i\}}}^{\prime} \frac{N!}{n_1! n_2!\cdots n_{m^2}!} \mathrm{e}^{-\beta \sum_{i=1}^{m^2}\varepsilon_i n_i},
\end{eqnarray}
where ${\sum}^{\prime}$ denotes that the indices $n_i$ in the sum satisfies the constraint $N=\sum_{i} \big<n_{i}\big>$. The factor ${n_1! n_2!\cdots n_{m^2}!}$ inside the sum arises from the fact that the packets are distinguishable from each other.
Our interest here lies only in calculating average quantities like the mean occupation number $\big<n_i\big>$ . These can be derived using either $Z_N$, or a normalized partition function ${Z^{\prime}_N}$ defined as
\begin{eqnarray}
{Z^{\prime}_N}=\frac{Z_N}{N!}={\sum_{\{n_i\}}}^{\prime} \frac{\mathrm{e}^{-\beta \sum_{i=1}^{m^2}\varepsilon_i n_i}} {n_1!  n_2!\cdots n_{m^2}!}.
\end{eqnarray}
The grand canonical partition function \cite{huang}, using ${Z^{\prime}_N}$,  is  given by
\begin{eqnarray}
\Xi=\sum_{N=0}^{\infty} \mathrm{e}^{\beta\mu N} {Z^{\prime}_N}.
\end{eqnarray}
For a large number of packets, the grand-canonical partition function
using either $Z^{\prime}_N$ or $Z_N$ will give the same predictions for
average quantities like $\big<n_{i}\big>$. After using Eq. $(7)$ and the
constraint $N=\sum_{i} \big<n_{i}\big>$ in Eq. $(8)$, the expression for
the grand-canonical partition function, after the usual algebra, can be written as
\begin{eqnarray}
\Xi= \prod_{i=1}^{m^2} \Bigg(\sum_{n_i=1}^{w_i}\frac{1}{n_i!} {\mathrm{e}^{-\beta (\varepsilon_i-\mu) n_i}}\Bigg),
\end{eqnarray}
where the capacity $w_i$ for the original lattice is given by the evolution equation $(1)$. From the grand-canonical partition function, the mean occupation numbers \cite{huang} of packets in a site $i$ of the network is calculated by using the relation
\begin{eqnarray}
\big<n_i\big>=-\frac{1}{\beta} \frac{\partial ln \Xi} {\partial \varepsilon_i}.
\end{eqnarray}
Using equation $(9)$ for $\Xi$ in the above equation, the following expression is obtained for mean number of packets in the site $i$
\begin{eqnarray}
\big<n_i\big>=\frac{\sum_{n=1}^{w_i}\frac{1}{(n-1)!} {\mathrm{e}^{-\beta (\varepsilon_i-\mu) n}}} {\sum_{n=1}^{w_i}\frac{1}{n!} {\mathrm{e}^{-\beta (\varepsilon_i-\mu) n}}}.
\end{eqnarray}
After substituting the value of $\varepsilon_i$ from Eq. $(3)$ in the
above expression, the mean number of packets in a site $i$  as a
function of its capacity $w_i$ is,
\begin{eqnarray}
\big<n_i\big>=\frac {\sum_{n=1}^{w_i}\frac{1}{(n-1)!} {(Aw_i})^n} {\sum_{n=0}^{w_i}\frac{1}{n!} {(Aw_i})^n},
\end{eqnarray}
where the parameter $A=c\mathrm{e}^{\beta\mu}$ is a function of the chemical potential $\mu$. The index $i$ here runs over all the sites of the network. In the limit of $\big< n_i\big> \rightarrow w_i$, the nodes with high capacity tend to become fully occupied. 
For large networks, and for large capacities $w_i\rightarrow \infty$, the above expression takes the form
\begin{eqnarray}
\big<n_i \big> \simeq (Aw_i)\frac{\partial}{\partial(Aw_i)} ln \bigg(\sum_{n=0}^{\infty}\frac{(Aw_i)^n}{n!} \bigg),
\end{eqnarray}
which simplifies in the large capacity limit to be
\begin{eqnarray}
\big<n_i \big> \simeq Aw_i.
\end{eqnarray}

Further from Eq. $(5)$ for the total number of packets $N$ present in the network, and Eq. $(13)$, an expression using which the chemical potential $\mu$ or $A$ can be determined, is obtained as follows
\begin{eqnarray}
N=\sum_i \frac {\sum_{n=1}^{w_i}\frac{1}{(n-1)!} {(Aw_i})^n} {\sum_{n=0}^{w_i}\frac{1}{n!} {(Aw_i})^n},
\end{eqnarray}
here the sum is over all the sites in the network. If $\rho=N/L$ be the occupation density where $L=M^2$ is the total number of sites in the network, and $P_w$ be the capacity distribution of the network, a relation to determine the parameter $A$ from Eq. $(15)$ (after taking the sum over capacities instead of over sites) can be written as follows
\begin{eqnarray}
\sum_w P_w \frac {\sum_{n=1}^{w}\frac{1}{(n-1)!} {(Aw})^n} {\sum_{n=0}^{w_i}\frac{1}{n!} {(Aw})^n}=\rho.
\end{eqnarray}
Eqs. $(13)$, and $(14)$ predict the mean occupation number $\big<n_i \big>$ in the branching hierarchical networks.
Since the capacities of sites in the tree-like hierarchical $2 D$ network tend to larger values as the depth of network increases, taking the limit of Eq. $(16)$ for large network size and large capacities, the above expression takes the form
\begin{eqnarray}
\sum_w P_w(Aw)\simeq \rho.
\end{eqnarray}
For a network where the capacity distribution $P_w$ is a power law $i.$$e.$ $P_w=cw^{-\alpha}$, where $\alpha$ is the exponent of the power law, Eq. $(17)$ can be used to obtain the value of $A$ to be:
\begin{eqnarray}
A \simeq (2-\alpha)\frac{\rho}{c}.
\end{eqnarray}
We note that the original lattice, and the strategy-I and strategy-II versions discussed in section II have capacity distributions of the power law form \cite{ajay}.

\subsection{Simulation results}

\begin{figure}[htb]
\begin{center}
\begin{tabular}{cc}
\includegraphics[height=5cm,width=6.5cm]{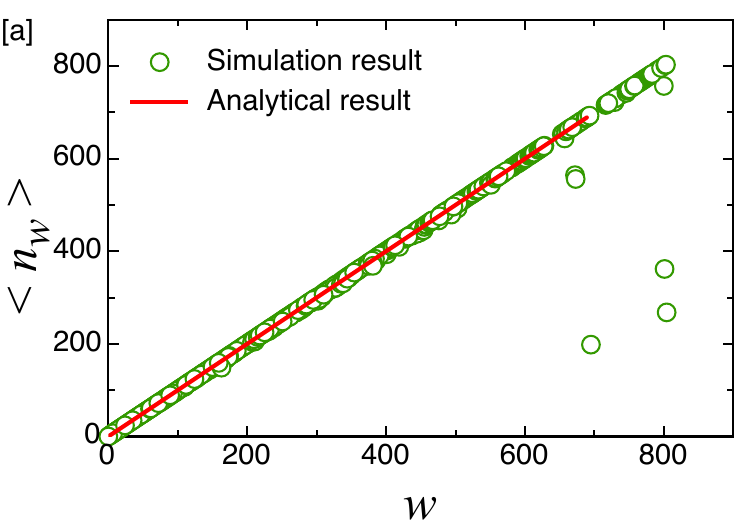}&
\includegraphics[height=5cm,width=6.5cm]{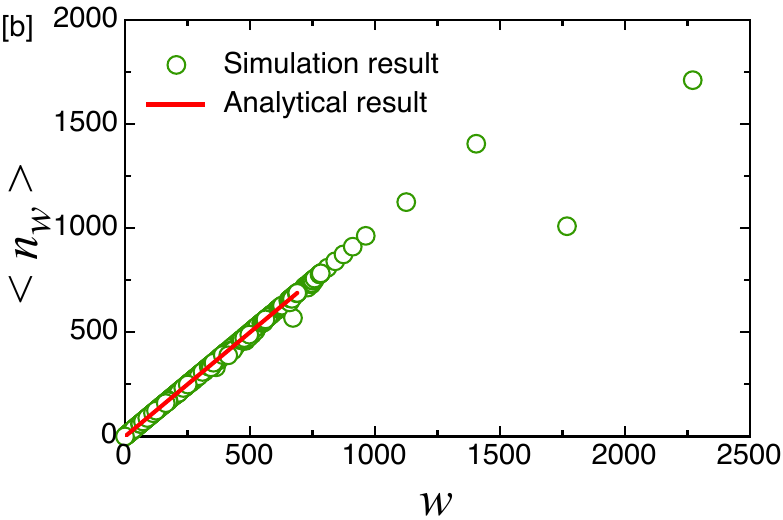}\\
\includegraphics[height=5cm,width=6.5cm]{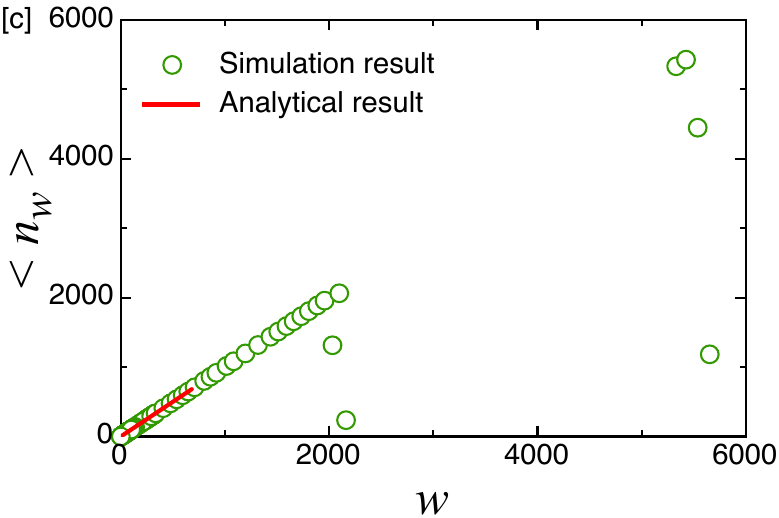}&
\includegraphics[height=5cm,width=6.5cm]{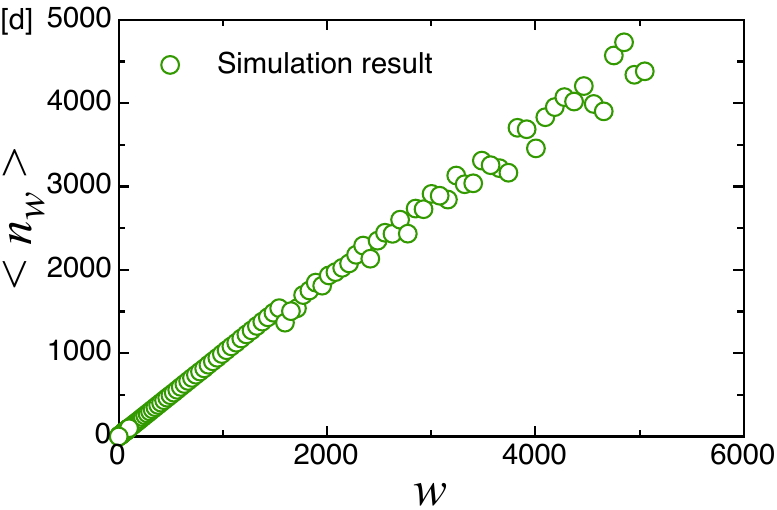}\\
\end{tabular}{}
\caption{\label{fig:mean} (Color online) The mean occupation number $\big<n_w\big>$ of sites with capacity $w$ of (a) the original lattice, (b) the strategy-I lattice, (c) the strategy-II lattice, and (d) the $V-$ lattice, averaged over $200$ realizations of networks of $L=100\times 100$ sites, corresponding to the packet density $\rho=3$.}
\end{center}
\end{figure}
We now simulate the transmission of information packets on the network. This transmission of information is modeled by packets of information hopping on the sites of the branching, hierarchical, heterogeneous, networks.
These packets could be the data packets of information in the Internet, which links computers of heterogeneous and high capacities capable of transmitting packets at high rates. For our networks, the packets are deposited at a randomly chosen site on the topmost layer of the network. Each site retains the number of packets which saturates its capacity and the remaining packets are transmitted further. A packet at a given site sees the nearest neighbor sites linked to itself. If the targeted neighboring site is not fully occupied (i.e. it has not saturated its capacity), the packet moves there and looks for the next site that has spare capacity. If the target site is fully occupied then the packet stops on the site which it occupies, and the transmission of the packet ends at that site. 
In this fashion, all packets hop from one site to another according to the vacancy available on the neighboring sites, and this process continues till all the packets come to rest at some site.
One realization of packet transmission ends here. The same number of packets is again placed on a randomly chosen site in the first layer of the hierarchical network, and the second realization of packet transmission proceeds. This stochastic process of packet transmission is repeated for a large number of realizations. Since our analytical expression for mean occupation number has the property that two nodes with the same capacity have the same occupation number, the mean occupation number is computed for each capacity instead of each site, by averaging over all the realizations of packet transmission.
We finally look at the average occupation number $\big<n_i \big>$ of the sites with the same capacity on the lattice, averaged over all the realizations, and compare it with the analytic result obtained in the earlier part of this section.

It can be seen from Fig. \ref{fig:mean} that the theoretical results match very well with the simulation results. Panels \ref{fig:mean}(a), \ref{fig:mean}(b), and \ref{fig:mean}(c) corresponding to the original lattice, the strategy-I, and the strategy-II lattices respectively show a very good agreement between the simulation and the theoretical results obtained \cite{footn}. However, the analytic form for the capacity distribution for the $V-$ lattice is unknown, so the theoretical match to it could not be shown in panel \ref{fig:mean}(d). Eq. $(13)$, along with the Eq. $(14)$ predicts the mean occupation number $\big<n_i \big>$ or statistical dynamics of sites on the branching hierarchical networks. This match between the theoretical and simulation predictions for the original lattice network and its different network topologies supports our finding that the dynamics of the transmission or flow of the packets of information or particles in weight bearing hierarchical networks follow the Maxwell-Boltzmann statistics.

\section{Percolation on the network}

We study site percolation on the hierarchical networks set up here due to its relevance for the problem of information transfer on the lattice. The site percolation problem is set up as follows:
If at any time, information packets occupy/do not occupy,  a given site in the network, that site is said to be occupied/unoccupied  at that time. The network then can be thought of being comprised of two sub-lattice networks, one being the unoccupied or free sub-lattice network, and the other being the occupied sub-lattice network. Here, we are interested in studying the transition to percolation in different lattice networks. From site percolation theory it is known that if the sub-lattice network of free or unoccupied sites percolates, it implies that a finite fraction of remaining unoccupied sites form a single connected giant cluster. We can identify this case as the state in which the single giant cluster could make communication possible throughout the network. If the free sub-lattice network does not percolate, the free or unoccupied sites then form mutually disconnected, and fragmented small clusters instead of a single giant cluster. In this state, communication through the network is not possible.
\begin{figure}[htb]
\begin{center}
\begin{tabular}{cc}
\includegraphics[height=5cm,width=6.5cm]{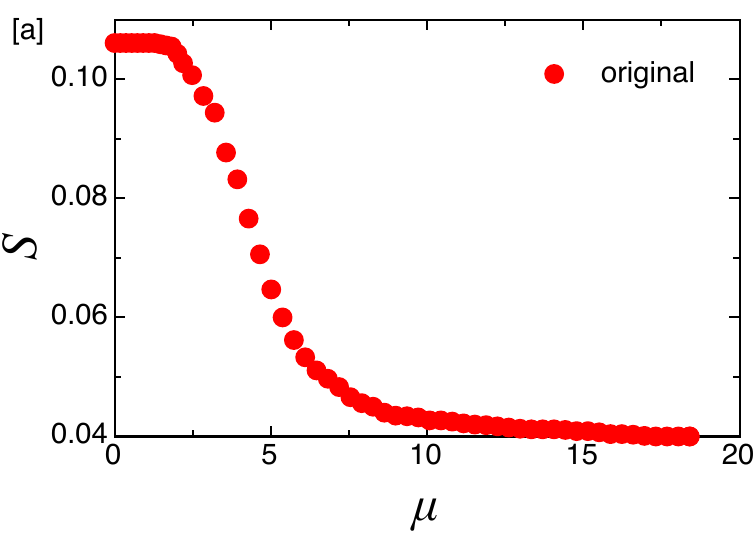}&
\includegraphics[height=5cm,width=6.5cm]{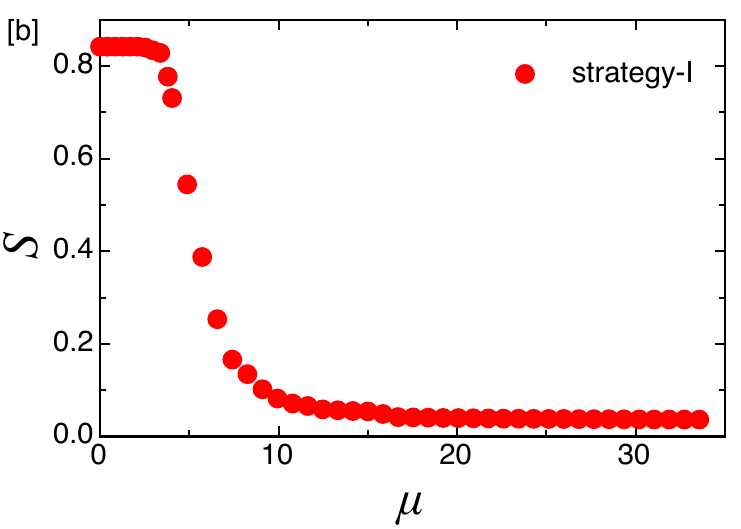}\\
\includegraphics[height=5cm,width=6.5cm]{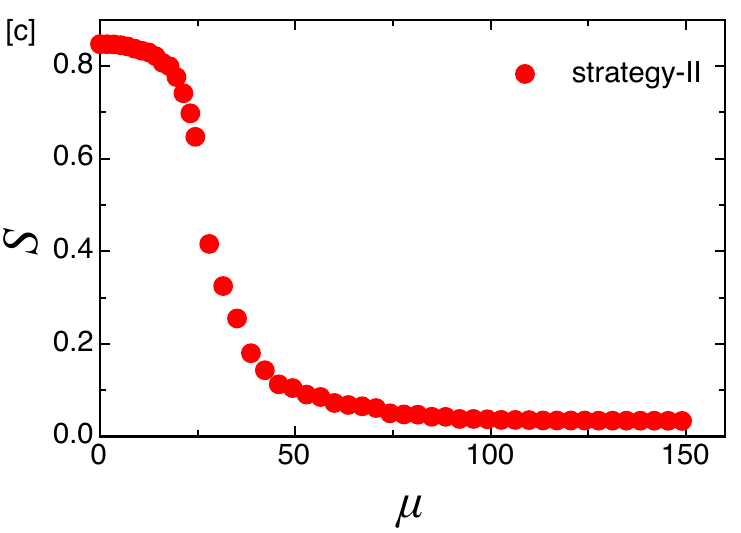}&
\includegraphics[height=5cm,width=6.5cm]{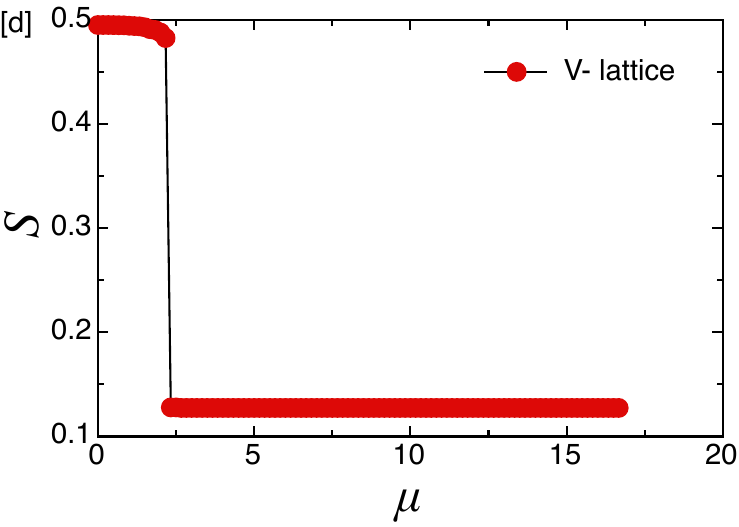}\\
\end{tabular}{}
\caption{\label{fig:percolS} (Color online) The percolation strength $S$ as a function of packet density $\mu$, averaged over $500$ realizations, for the unoccupied sub-lattice in (a) the original lattice, (b) the strategy - I lattice, (c) the strategy - II lattice, and (d) the $V-$ lattice networks of $L=100\times 100$ sites.}
\end{center}
\end{figure}
\begin{figure}[htb]
\begin{center}
\begin{tabular}{cc}
\includegraphics[height=5cm,width=6.5cm]{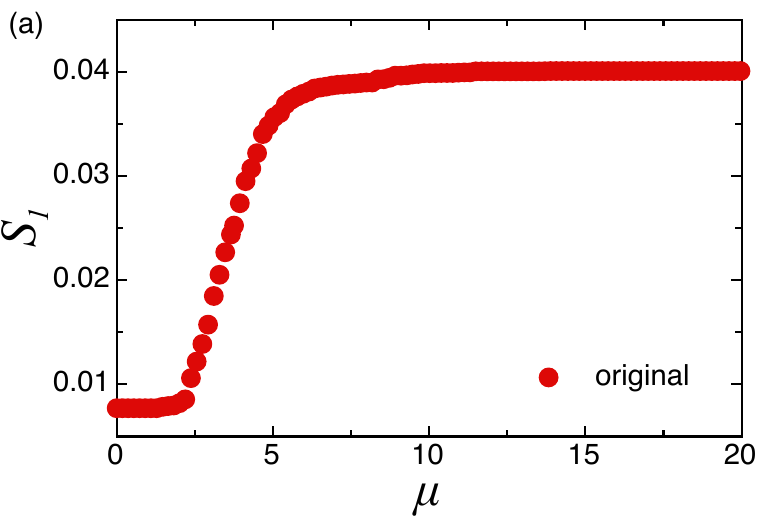}&
\includegraphics[height=5cm,width=6.5cm]{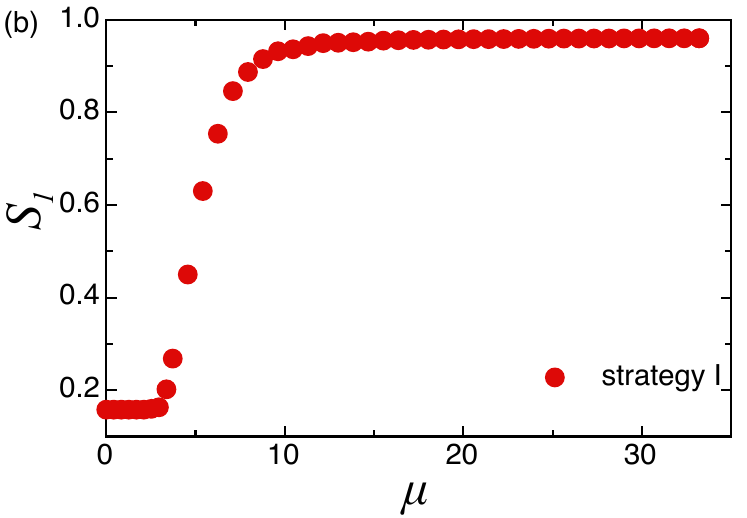}\\
\includegraphics[height=5cm,width=6.5cm]{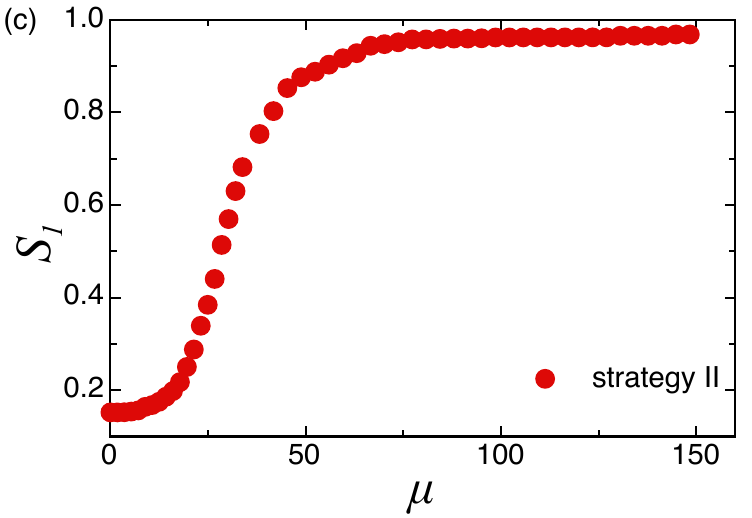}&
\includegraphics[height=5cm,width=6.5cm]{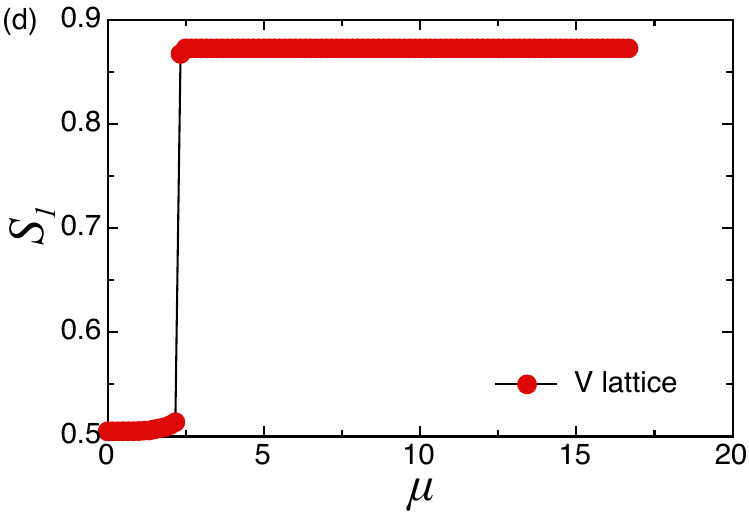}\\
\end{tabular}{}
\caption{\label{fig:percol} (Color online) The percolation strength $S_1$ as a function of packet density $\mu$, averaged over $500$ realizations, for the occupied sub-lattice in (a) the original lattice, (b) the strategy - I lattice, (c) the strategy - II lattice, and (d) the $V-$ lattice networks of $L=100\times 100$ sites.}
\end{center}
\end{figure}

It is interesting to study the transition to percolation in the occupied or unoccupied sub-lattice network. We anticipate a transition from the non-percolating to the percolating state when the packet density $\mu$ changes. In order to analyze the transition from the non-percolating to the percolating state in networks, we simulate the stochastic dynamics of packet transmission in the original lattice, the strategy-I and the strategy-II lattices, and the $V-$ lattice for different values of the packet density $\mu$. Two types of transition are seen in the system, as we will see in the next section.

\subsection{The two transitions: Percolation and explosive percolation} 

To study the phase transition, we define the order parameter to be the percolation strength $S=S_m/L$ \cite{radicchi}, where $S_m$ indicates the number of sites belonging to the largest connected cluster of the unoccupied sub-lattice networks, and $L=M^2$, the total number of sites in the network. The parameter percolation strength $S$ of the unoccupied sub-lattice corresponding to the original, the strategy -I, the strategy -II lattices and the $V-$ lattice is shown in panels \ref{fig:percolS}(a), \ref{fig:percolS}(b), \ref{fig:percolS}(c) and \ref{fig:percolS}(d). This order parameter satisfies scaling relations for quantities like the susceptibility and the critical packet density, as we will see in the next section. Scaling relations of the type seen in Ref. \cite{radicchi} are seen for the percolation strength $S_1=(1-S)$, which is a measure of the fraction of occupied sites or of the size of the occupied sub-lattice network as a function of the packet density $\mu$ averaged over a number of realizations as shown in Fig. \ref{fig:percol}. The panels \ref{fig:percol}(a), \ref{fig:percol}(b), \ref{fig:percol}(c), and \ref{fig:percol}(d) display the percolation strength $S_1$ as a function of the packet density $\mu$ for the original, the strategy-I, the strategy-II lattices, and the V-lattice respectively. The first three panels \ref{fig:percol}(a), \ref{fig:percol}(b), \ref{fig:percol}(c) corresponding to the original, the strategy-I, and the strategy-II lattice network demonstrate that the transition from non-percolating to a percolating state in these lattices is actually a continuous phase transition. In each panel of Fig. \ref{fig:percol}, $\mu_c$ is the critical packet density where a transition from a non-percolating to a percolating state occurs.
\begin{figure}[htb]
\begin{center}
\begin{tabular}{ccc}
\includegraphics[height=5cm,width=5cm]{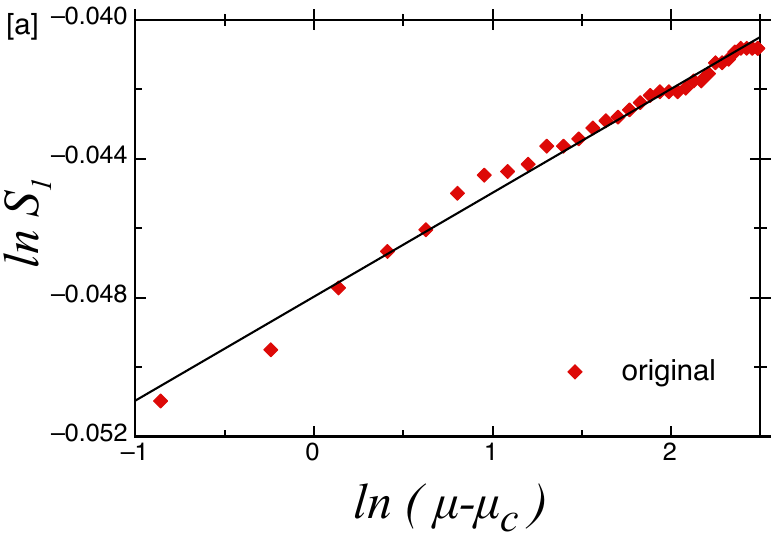}&
\includegraphics[height=5cm,width=5cm]{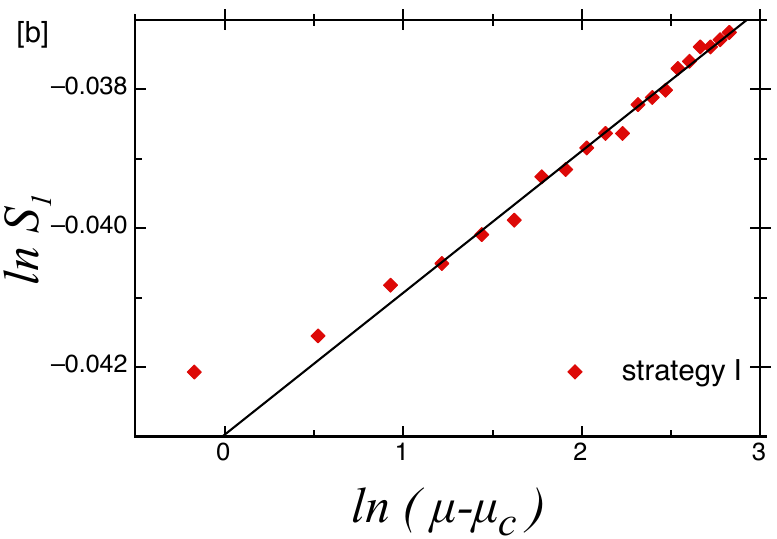}&
\includegraphics[height=5cm,width=5cm]{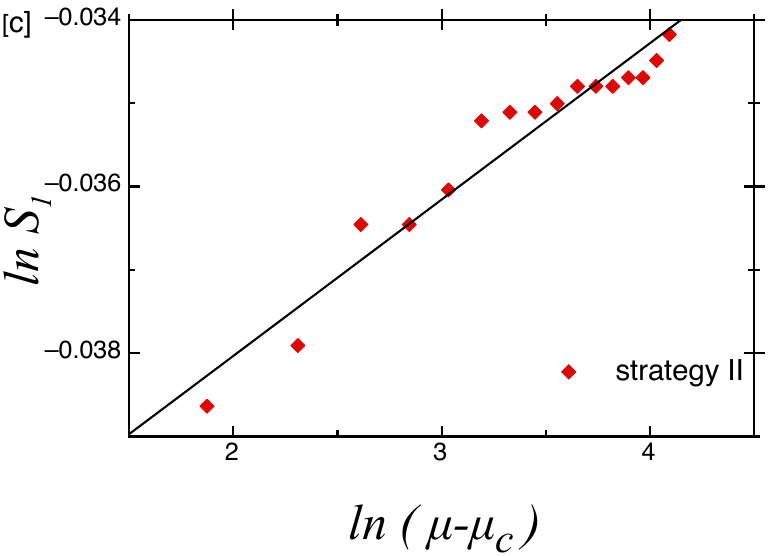}\\
\end{tabular}{}
\caption{\label{fig:crit_perc} (Color online) The percolation strength $S_1$ as a function of packet density $\mu$, displays power law behavior $S_1 \propto (\mu-\mu_c)^{\eta}$ corresponding to $500$ realizations, for the occupied sub-lattice in (a) the original lattice with $\eta=0.0161\pm0.0067$, (b) the strategy - I with $\eta=0.0204\pm0.0047$, and (c) the strategy - II lattice with $\eta=0.0187\pm0.0257$, networks of $L=100\times 100$ sites.}
\end{center}
\end{figure}

The most striking result is however observed for the critical case of the original lattice the $V-$ lattice \cite{ajay}, in which a sharp, discontinuous percolation transition appears unlike the other three cases.

\subsection{Finite size scaling and critical exponents:}

The percolation strength $S_1$, which is a measure of the fraction  of occupied sites when plotted versus $(\mu-\mu_c)$, shows that $S_1$ displays power law behavior of the form $S_1 \sim (\mu-\mu_c)^{\eta}$ for the all the three lattice network indicating the existence of a continuous percolation transition as shown in Fig. \ref{fig:crit_perc}, except for the case of the critical $V-$ lattice. For the $V-$ lattice, the percolation transition is discontinuous, and hence scaling does not work. The exponents $\eta$ of the power law for the original, the strategy-I, and the strategy-II lattices are found to be $0.0161$, $0.0204$, and $0.0187$ respectively.
\begin{figure}[htb]
\begin{center}
\begin{tabular}{ccc}
\includegraphics[height=5cm,width=5cm]{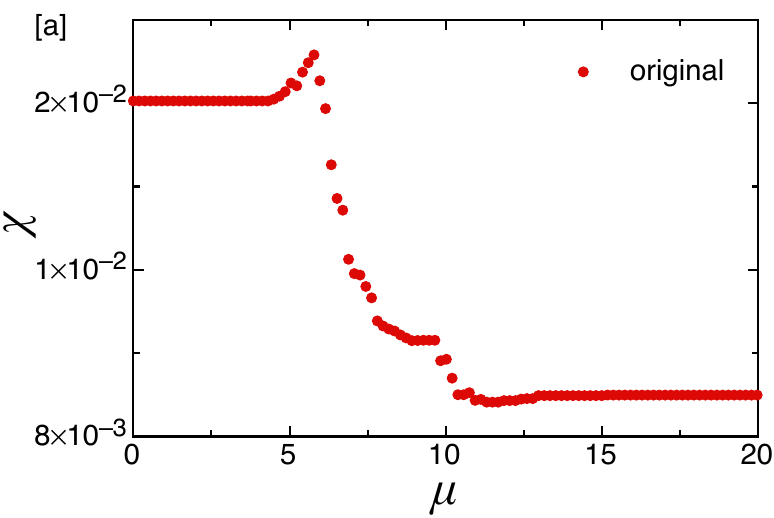}&
\includegraphics[height=5cm,width=5cm]{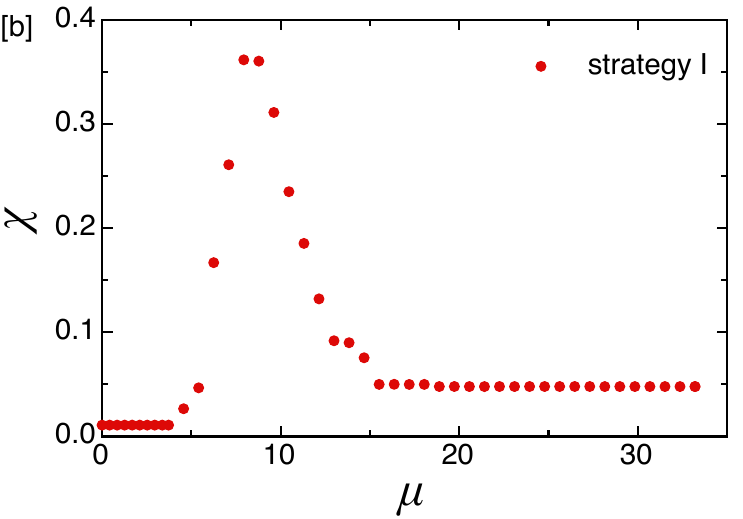}\\
\includegraphics[height=5cm,width=5cm]{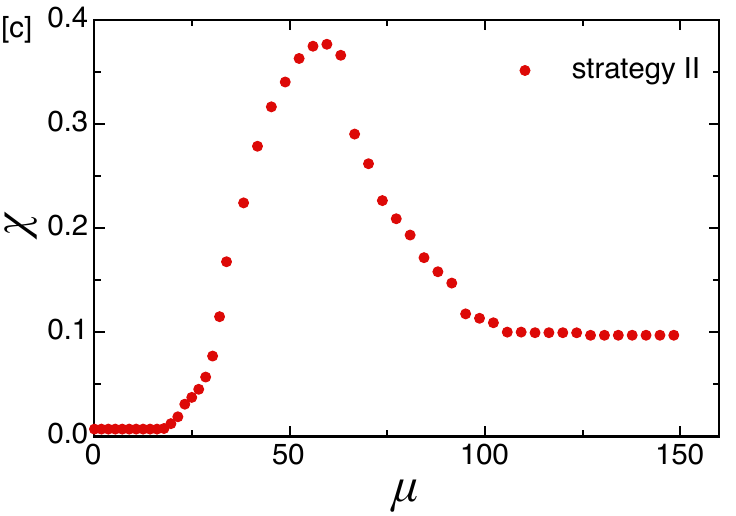}&
\includegraphics[height=5cm,width=5cm]{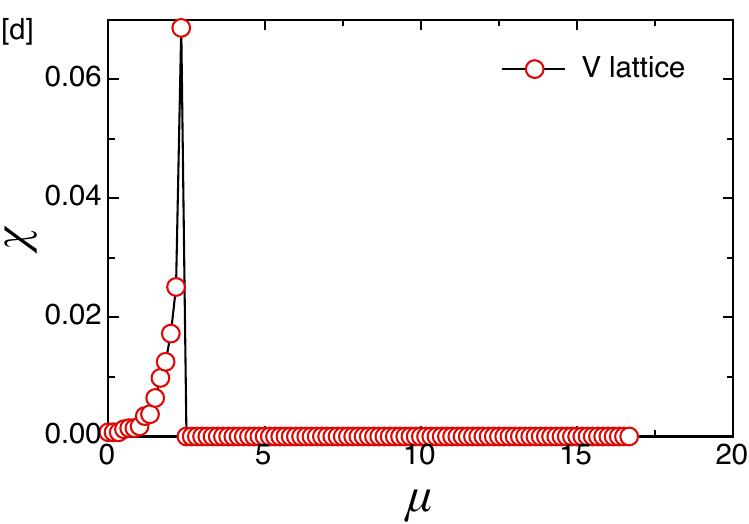}\\
\end{tabular}{}
\caption{\label{fig:perco_sd} (Color online) The susceptibility $\chi$ as a function of packet density $\mu$, averaged over $500$ realizations, for the occupied sub-lattice in (a) the original lattice, (b) the strategy - I lattice, (c) the strategy -II lattice and (d) $V-$ lattice networks of $L=100\times 100$ sites.}
\end{center}
\end{figure}
\begin{figure}[htb]
\begin{center}
\begin{tabular}{ccc}
\includegraphics[height=5cm,width=5cm]{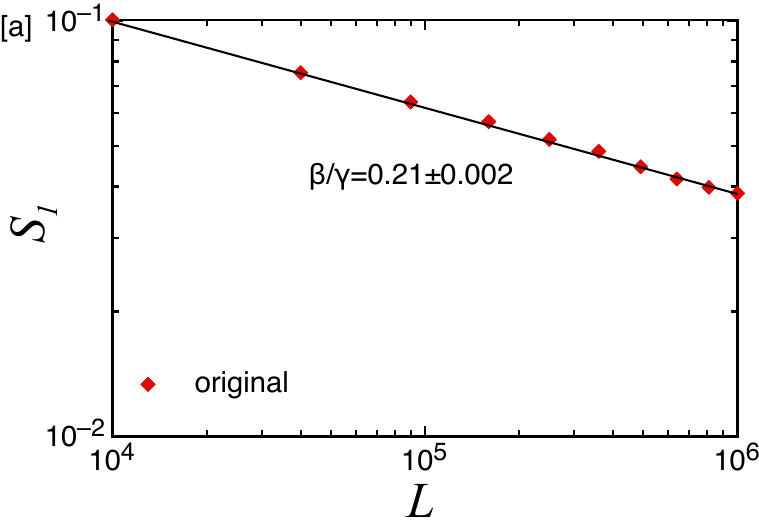}&
\includegraphics[height=5cm,width=5cm]{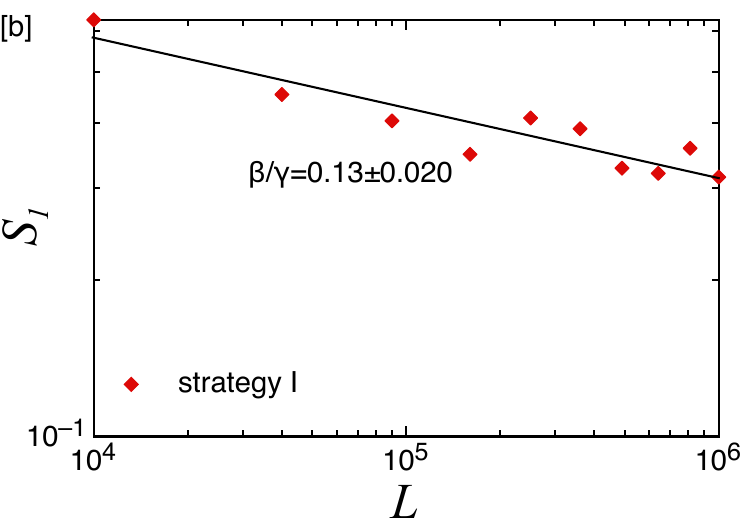}&
\includegraphics[height=5cm,width=5cm]{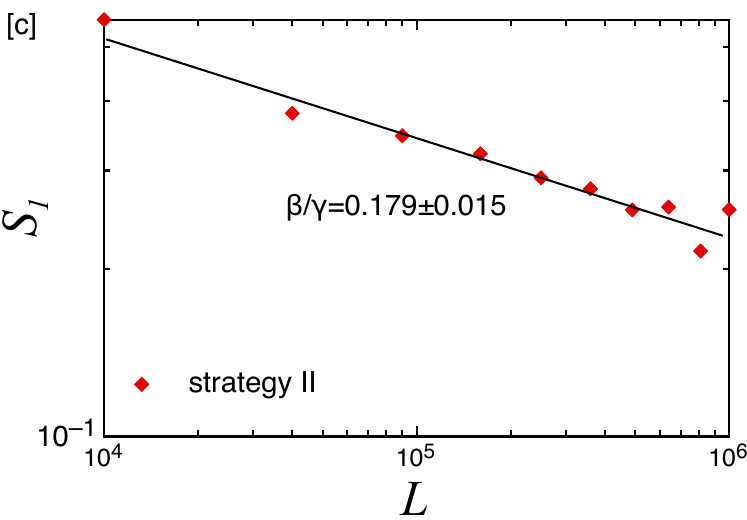}\\
\end{tabular}{}
\caption{\label{fig:ps_scaling} (Color online) The percolation strength $S_1$ at the critical point $\mu_c$, as a function of the total number of sites $L$ and corresponding to $500$ realizations, displays power law behavior $S_1 \sim L^{-\beta/\nu}$, for the occupied sub-lattice in (a) the original lattice with $\beta/\nu=0.2101\pm 0.0025$, (b) the strategy - I with $\beta/\nu=0.1306\pm 0.0202$, and (c) the strategy - II networks with $\beta/\nu=0.1798\pm 0.0148$.}
\end{center}
\end{figure}
\begin{figure}[htb]
\begin{center}
\begin{tabular}{ccc}
\includegraphics[height=5cm,width=5cm]{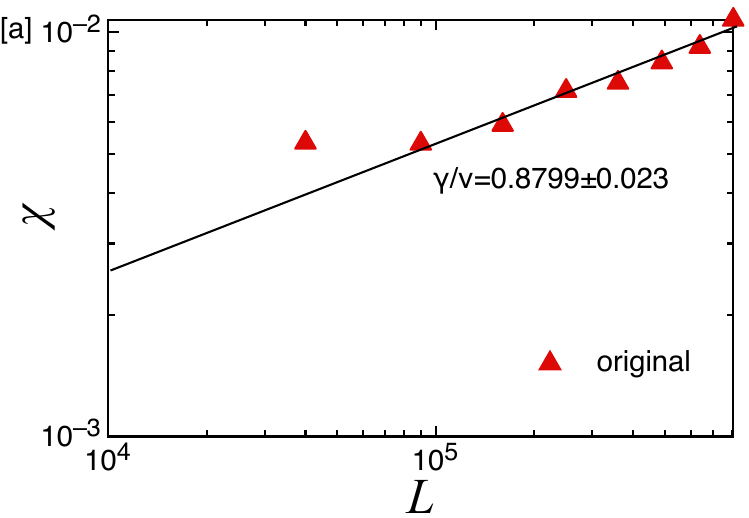}&
\includegraphics[height=5cm,width=5cm]{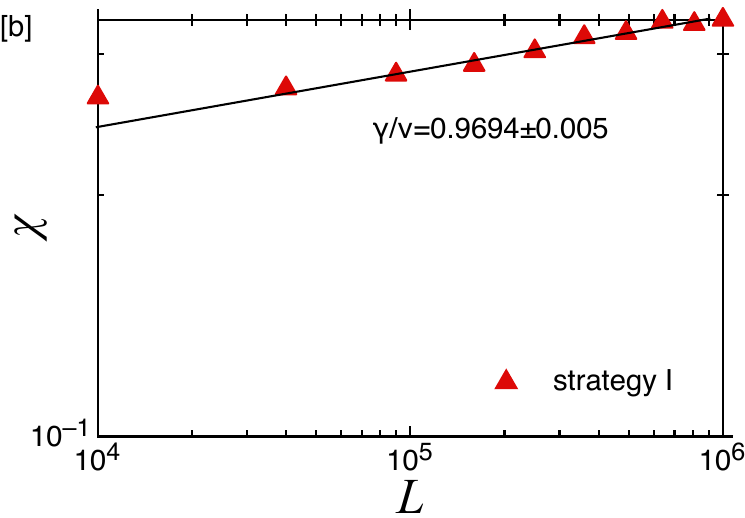}&
\includegraphics[height=5cm,width=5cm]{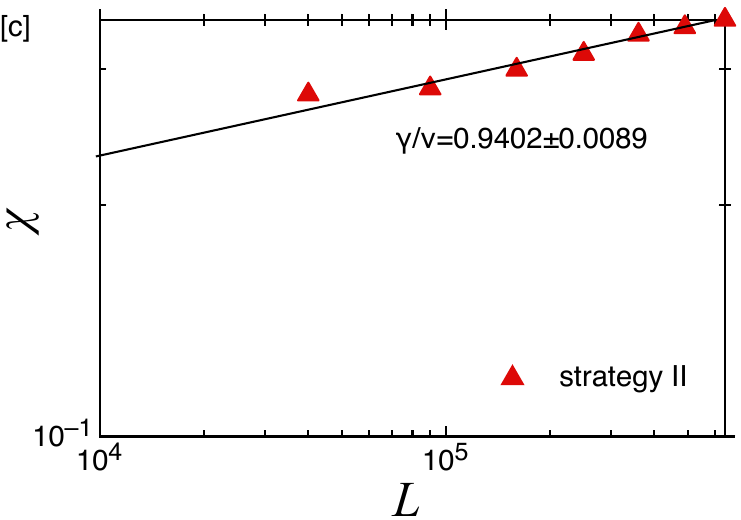}\\
\end{tabular}{}
\caption{\label{fig:sc_scaling} (Color online) The susceptibility $\chi$ at the critical point $\mu_c$, as a function of the total number of sites $L$ and corresponding to $500$ realizations, displays power law behavior $\chi \sim L^{\gamma/\nu}$, for the occupied sub-lattice in (a) the original lattice with $\gamma/\nu=0.8799\pm 0.0231$, (b) the strategy - I with $\gamma/\nu=0.9694\pm 0.0053$, and (c) the strategy - II networks with $\gamma/\nu=0.9402\pm 0.0088$.}
\end{center}
\end{figure}
\begin{figure}[htb]
\begin{center}
\begin{tabular}{ccc}
\includegraphics[height=5cm,width=5cm]{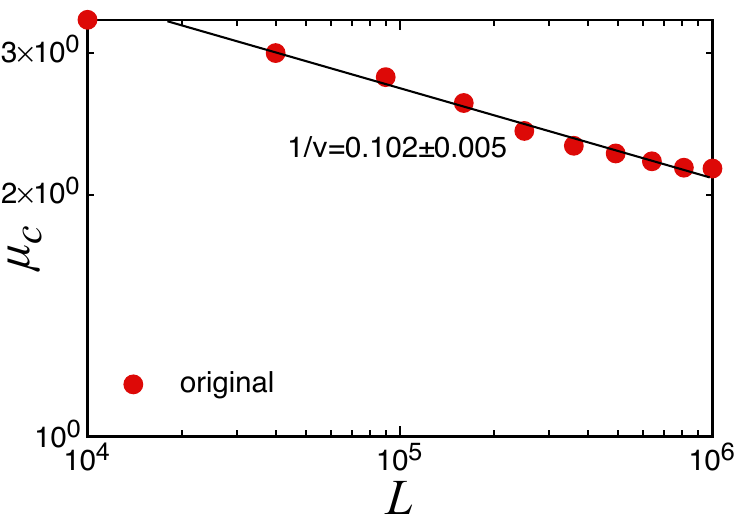}&
\includegraphics[height=5cm,width=5cm]{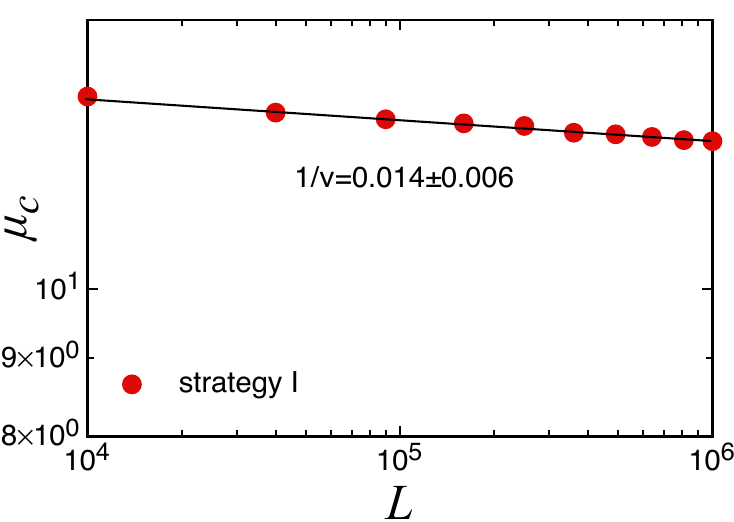}&
\includegraphics[height=5cm,width=5cm]{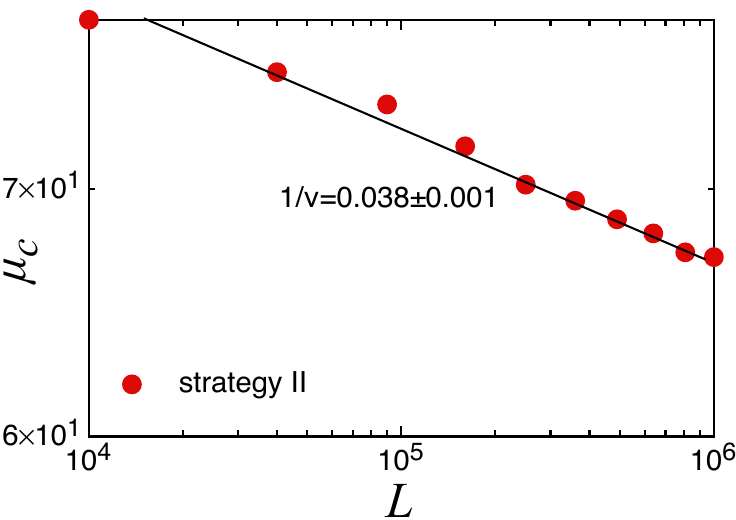}\\
\end{tabular}{}
\caption{\label{fig:explo_critd} (Color online) The log - log plot of critical packet density $\mu_c$ as a function of the network of $L$ sites for (a) the original lattice with $1/\nu=0.1017\pm0.0049$; $d=2.1568$, (b) the strategy -I lattice with $1/\nu=0.0138\pm0.006$; $d=2.7156$, (c) the strategy-II lattice with $1/\nu=0.0378\pm0.0012$; $d=4.7266$.}
\end{center}
\end{figure}

For continuous percolation transitions, the percolation strength $S_1$ of a network of $L$ sites, obeys the following finite size scaling (FSS) relation
\begin{equation}
S_1=L^{-\beta/\nu}f[(\mu-\mu_c)L^{1/\nu}],
\end{equation}
where $\beta$ is the critical exponent for the order parameter, $\nu$ that of the correlation length, and $f(\cdot)$ is a universal function.
We also define the susceptibility $\chi=L\sqrt{\big<S_1^2\big>-{\big<S_1\big>}^2}$, which quantifies the amplitude of the fluctuations of the percolation strength $S_1$. The susceptibility in our cases, shows finite size scaling, with the scaling function
\begin{equation}
\chi=L^{\gamma/\nu}g[(\mu-\mu_c)L^{1/\nu}],
\end{equation}
where $\gamma$ is a critical exponent of the susceptibility and $g(\cdot)$ a universal function. The susceptibility $\chi$ is used to determine the critical point $\mu_c$. The panels \ref{fig:perco_sd}(a), \ref{fig:perco_sd}(b), \ref{fig:perco_sd}(c) and \ref{fig:perco_sd}(d) display the susceptibility $\chi$ against the packet density $\mu$ for the original, the strategy-I, the strategy-II lattices and the $V-$ lattice respectively. The critical point $\mu_c$ in these cases is determined from the absolute maximum of the susceptibility $\chi$ \cite{radicchi} as shown in panels \ref{fig:perco_sd}(a), \ref{fig:perco_sd}(b), and \ref{fig:perco_sd}(c).

It is known that \cite{radicchi}, the susceptibility, and the percolation strength at critical point are directly related, and their critical exponents satisfy the equality $\beta/\nu+\gamma/\nu=1$. For the original, the strategy-I and the strategy-II lattices, where a continuous phase transition is observed, we computed the critical exponents by examining the finite size scaling of the two variables, the susceptibility $\chi$ and the percolation strength $S_1$ at the critical point and comparing these with Eqs. $(19)$ and $(20)$ (see  the panels of Fig. \ref{fig:ps_scaling} and Fig. \ref{fig:sc_scaling}). From the panels \ref{fig:ps_scaling}(a) and  \ref{fig:sc_scaling}(a) for the original lattice, it can be seen that the values of exponents $\beta/\nu$ and $\gamma/\nu$ fall between $0$ and $1$, and their sum $\beta/\nu$ + $\gamma/\nu$ is very close to $1$. The same can be observed for the strategy-I from the panels \ref{fig:ps_scaling}(b) and \ref{fig:sc_scaling}(b), and for the strategy-II lattice from the panels \ref{fig:ps_scaling}(c) and \ref{fig:sc_scaling}(c). If the transition is discontinuous as in the case of the V- lattice, the finite size scaling law from Eq. (19) does not hold, and the percolation transition is of the explosive type. Since for the V-lattice, $S_1$ at critical point does not vanish in the infinite size limit, the ratio $\beta/\nu=0$, and the ratio $\gamma/\nu=1$ as the susceptibility $\chi$ is extensive (from panel \ref{fig:perco_sd}(d)) \cite{radicchi}. 
Thus the relation between the critical exponents is satisfied for this case as well.

It is also observed that the critical packet density $\mu_c(L)$ corresponding to a network of $L$ sites obeys the relation
\begin{equation}
\mu_{c}(L)=\mu_c+dL^{-1/\nu},
\end{equation}
as can be verified from the panels \ref{fig:explo_critd}(a), \ref{fig:explo_critd}(b), \ref{fig:explo_critd}(c) for the original, the strategy-I, and strategy-II lattice networks, respectively. The parameter $d$ is a constant in this expression. This expression can be used to compute the values of $\nu$ corresponding to the three lattices.
The values of all three exponents $\beta$, $\gamma$ and $\nu$ are tabulated in Table I for the original, the strategy-I and the strategy-II lattices.

\begin{figure}[htb]
\begin{center}
\includegraphics[height=5.5cm,width=6.5cm]{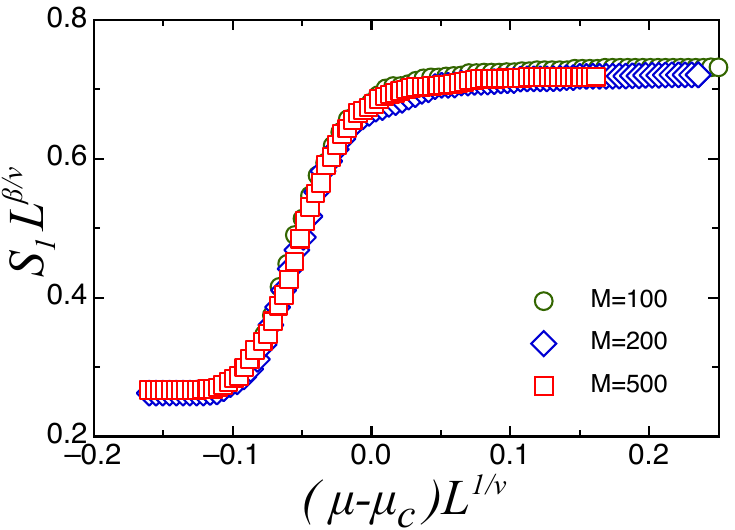}\\
\caption{\label{fig:ps_scaled} (Color online) Scaling of percolation strength $S_1$ corresponding to different lattice sizes $L$ $i.$ $e.$, $S_1L^{\beta/\nu}$ versus $(\mu-\mu_c)L^{1/\nu}$ for the original lattice with $\beta/\nu=0.224$, $1/\nu=0.11$.}
\end{center}
\end{figure}

\begin{table}[htbp]
\caption {The ratios of the critical exponents $\beta/\nu$, $\gamma/\nu$ and $1/\nu$ of the percolation transition for the occupied sub-lattice of the original, the strategy-I and the strategy-II lattices.}
\vskip0.2in
\begin{center}
    \begin{tabular}{ | c | c | c | c | c |}
     \hline
    Network & $\beta/\nu$ & $\gamma/\nu$ & $1/\nu$ & $\beta/\nu+\gamma/\nu \simeq1$ \\
    \hline
    Original & $0.2101\pm0.0025$ & $0.8799\pm0.0231$ & $0.1017\pm0.0049$ & $1.09$ \\
    \hline
    Strategy-I & $0.1306\pm0.0202$ & $0.9694\pm0.0053$ & $0.0138\pm0.006$ & $1.1$ \\
    \hline
    Strategy-II & $0.1798\pm0.0148$ & $0.9402\pm0.0088$ & $0.0378\pm0.0012$ & $1.12$ \\
    \hline
    \end{tabular}
\end{center}
\end{table}
The scaling behavior of the percolation strength $S_1$ for the occupied sub-lattice of the original lattice networks is shown in Fig. \ref{fig:ps_scaled}.

The finite-size scaling (FSS) relations of the percolation strength and the susceptibility for the continuous percolation transition in the unoccupied sub-lattice networks of the original, the strategy -I and strategy -II lattices would be $S=L^{\beta/\nu}f[(\mu-\mu_c)L^{-1/\nu}]$, and $\chi=L^{-\gamma/\nu}g[(\mu-\mu_c)L^{-1/\nu}]$, respectively.
The values of all three exponents $\beta$, $\gamma$ and $\nu$ corresponding to unoccuppied sub-lattice network are tabulated in Table II for the original, the strategy-I and the strategy-II lattices.

\noindent {\bf Table II:} The values of the critical exponents $\beta$, $\gamma$ and $\nu$ of the percolation transition for the unoccupied sub-lattice of the original, the strategy-I and the strategy-II lattices.
\vskip0.2in
\begin{center}
    \begin{tabular}{ | c | c | c | c | c |}
     \hline
    Network & $\beta$ & $\gamma$ & $-1/\nu$ & $\beta/\nu+\gamma/\nu \simeq1$ \\
    \hline
    Original & $2.19\pm0.0064$ & $0.86\pm0.0753$ & $0.36\pm0.13$ & $1.1002$ \\
    \hline
    Strategy-I & $2.27\pm0.0348$ & $0.256\pm0.0214$ & $0.43\pm0.02$ & $1.0922$ \\
    \hline
    Strategy-II & $2.265\pm0.0094$ & $0.114\pm0.0368$ & $0.437\pm0.026$ & $1.0463$ \\
    \hline
    \end{tabular}
\end{center}

\begin{figure}[htb]
\begin{center}
\begin{tabular}{cc}
\includegraphics[height=5cm,width=5cm]{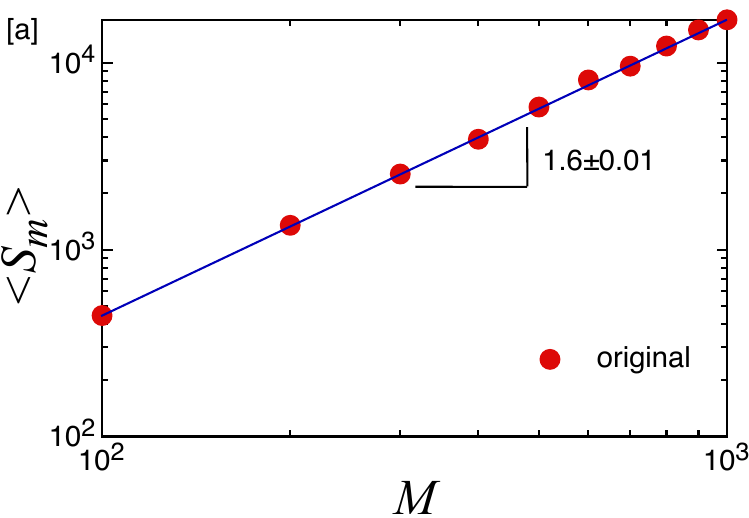}&
\includegraphics[height=5cm,width=5cm]{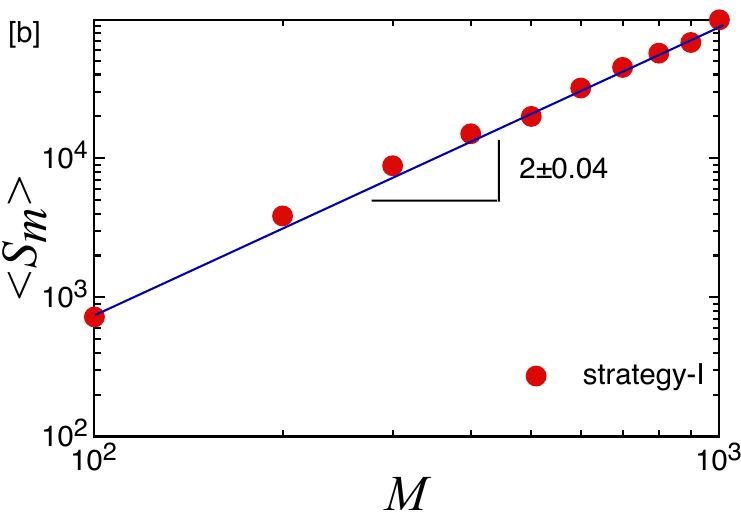}\\
\includegraphics[height=5cm,width=5cm]{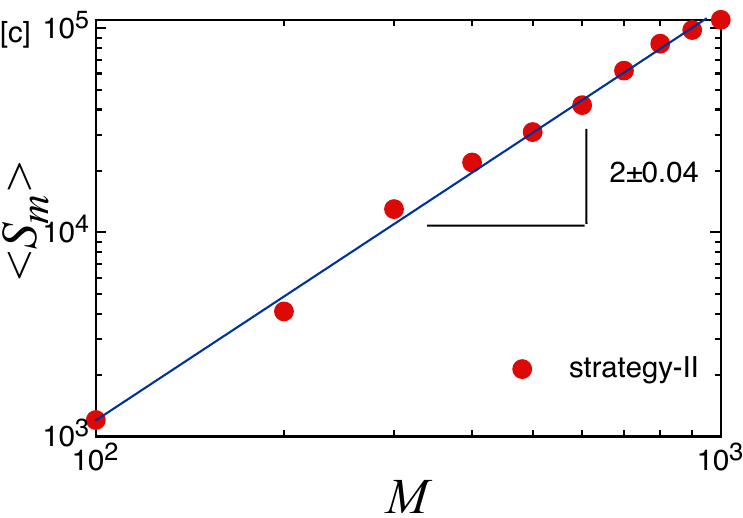}&
\includegraphics[height=5cm,width=5cm]{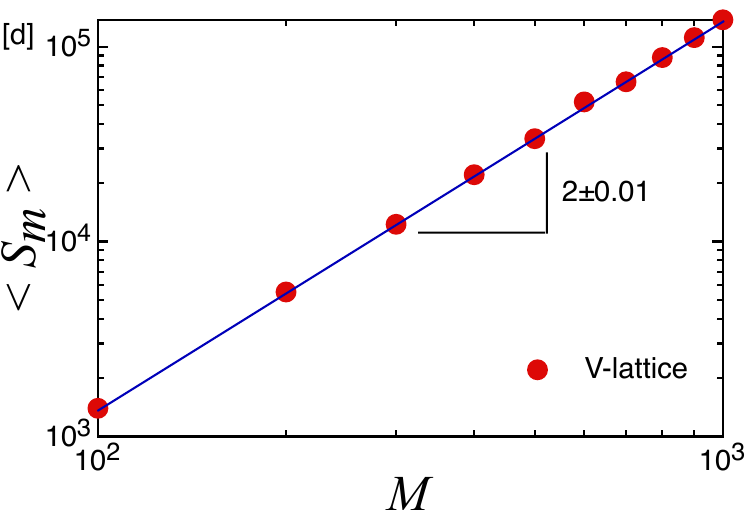}\\
\end{tabular}{}
\caption{\label{fig:explo} (Color online) The log - log plot of $S_m$, cluster size of the largest unoccupied sub-networks at the onset of percolation transition (at the jump in case of the V-lattice), as a function of the lattice side $M$ for (a) the original, (b) the strategy-I, (c) the strategy-II, and (d) the $V-$ lattice networks.}
\end{center}
\end{figure}

We, further computed the size of the largest cluster $S_m$ of unoccupied sub-lattice at the onset of the phase transition (at the jump in case of the V-lattice), as a function of the lattice side $M$. From the three panels of Fig. \ref{fig:explo} corresponding to three different topologies, the size of the largest cluster $S_m$ prior to the transition threshold does not grow proportionally to the lattice side $M$, instead it obeys a power law of the form of $S_m\sim M^{\zeta}$. We note that in the case of the original lattice $\zeta=1.6$, whereas for strategies I and II, $\zeta \approx 2$ indicating that the strategies have been successful in incorporating a substantial fraction of the sites of the lattice in the maximal cluster.

\section{Conclusions}

The study of the transport of packets on hierarchical networks has yielded some intriguing insights into the statistics and the percolation behavior of such systems. Most earlier studies of transport on networks have been performed on either scale-free or random networks and not on hierarchical ones. The models used by us to study traffic of packets, are typical example of tree- or river- like uni-directional branching hierarchical networks, as in computer networks with backbones in which the capacity distribution of sites is more significant than the degree distribution for the analysis of traffic. These networks are highly heterogeneous, and display power law behavior for the capacity distribution. If distinguishable packets are transported on these networks, an analytical expression for the mean occupation number of sites on a hierarchical network can be obtained as a function of its capacity distribution, which matches with the numerical simulations, and confirms that the packet distribution satisfies the Maxwell-Boltzmann distribution.

Site percolation studies on networks of a given geometry and capacity distribution can also yield important insights into their efficiency for information transport. Hence, we studied the site percolation problem for different topologies of the hierarchical network. We have seen that the percolation transition in the original model and its two modified versions is a continuous transition, whereas the percolation transition is explosive (discontinuous) in the critical case of the original lattice, i.e. the $V-$ lattice. We have recently seen that the avalanche distribution on the $V-$ lattice shows power law behavior, unlike the peaked behavior seen on the other lattices. Thus the $V-$ lattice is a special lattice, which promotes rapid transport of traffic. Recently, the explosive transition has been explored extensively in various scale-free and random networks. Some results indicate that the explosive percolation transition, for certain classes of networks, is not a genuine discontinuous transition, and the discontinuity is a consequence of the finite size of the lattice. However, these lattices are quite different from our hierarchical lattices, and the question of whether the explosive percolation on the $V-$ lattice will survive the infinite size limit is worth independent study. We hope to explore this question, as well as the question of the optimality of the $V-$ geometry for efficient transport in future work.

\section{Acknowledgments}
AD thanks the University Grants Commission, India, for a fellowship. NG thanks CSIR, India, for partial support.


\begin{thebibliography}{90}

\bibitem{strogatz} S. H. Strogatz, Nature (London) {\bf410}, 268 (2001). 
\bibitem{barabasi} R. Albert and A.-L. Barab\'{a}si, Rev. Mod. Phys. {\bf74}, 47 (2002). 
\bibitem{newman} M. E. J. Newman, SIAM Rev. {\bf45}, 167 (2003).
\bibitem{csabai} I. Csabai, J. Phys. A {\bf27}, L417 (1994).
\bibitem{takayasu} M. Takayasu, H. Takayasu, and T. Salo, Physica A {\bf233}, 824 (1996).
\bibitem{radner} R. Radner, Econometrica {\bf61}, 1109 (1993).
\bibitem{ohira} T. Ohira and R. Sawatari, Phys. Rev. E {\bf58}, 193 (1998).
\bibitem{trety} A. Tretyakov, H. Takayasu, and M. Takayasu, Physica A {\bf253}, 315 (1998).
\bibitem{guimera} R. Guimer\`{a}, A. Arenas, A. Diaz-Guilera, and F. Giralt, Phys. Rev. E {\bf66}, 026704 (2002).
\bibitem{brajendra} B. K. Singh and N. Gupte, Phys. Rev. E {\bf71}, 055103 (R) (2005).
\bibitem{mukherjee} S. Mukherjee and N. Gupte, Phys. Rev. E {\bf77}, 036121 (2008).
\bibitem{kahng} K. I. Goh, B. Kahng, and D. Kim, Phys. Rev. Lett. {\bf87}, 278701 (2001). 
\bibitem{goh} K. I. Goh et al., Proc. Natl. Acad. Sci. U.S.A. {\bf99}, 12583 (2002). 
\bibitem{goh2} K. I. Goh, B. Kahng, and D. Kim, Physica A {\bf318}, 72 (2003). 
\bibitem{barth} M. Barthelemy, Eur. Phys. J. B {\bf38}, 163 (2004).
\bibitem{ghim} C. M. Ghim et al., Eur. Phys. J. B {\bf38}, 193 (2004).
\bibitem{newman2} M. E. J. Newman, Soc. Networks {\bf27}, 39 (2005).
\bibitem{tadic} B. Tadi\'{c}, S. Thurner, and G. J. Rodgers, Phys. Rev. E {\bf69}, 036102 (2004).
\bibitem{manna} S. S. Manna and A. Chatterjee, Physica A {\bf390}, 177 (2011).
\bibitem{barth2} M. Barthelemy, B. Gondran, and E. Guichard, Phys. Rev. E {\bf66}, 056110 (2002). 
\bibitem{sole} S. Valverde and R. V. Sole, Physica A {\bf312}, 636 (2002); S. Valverde and R. V. Sole, Eur. Phys. J. B {\bf38}, 245 (2004).
\bibitem{monero} Y. Moreno, R. Pastor-Satorras, A. Vazquez, and A. Vespignani, Europhys. Lett. {\bf62}, 292 (2003).
\bibitem{holme} P. Holme, Adv. Complex Syst. {\bf6}, 163 (2003).
\bibitem{steger} J. Steger, P. Vaderna, and G. Vattay, Physics A {\bf360}, 134 (2006).
\bibitem{solla} I. Glauche, W. Krause, R. Sollacher, and M. Greiner, Physica A {\bf341}, 677 (2004).
\bibitem{zhao} L. Zhao, Y.-C. Lai, and K. Park, Phys. Rev. E {\bf71}, 026125 (2005).
\bibitem{moura} A. P. S. de Moura, Phys. Rev. E {\bf71}, 066114 (2005).
\bibitem{germano} R. Germano, and A. P. S. de Moura, Phys. Rev. E {\bf74}, 036117 (2006).
\bibitem{latora}V. Latora and M. Marchiori, Phys.Rev. Lett. {\bf 87}, 198701(2001).
\bibitem{basumohanty} M. Basu and P.K. Mohanty, J. Stat. Mech. 10014(2010).
\bibitem{janaki} T. M. Janaki, and N. Gupte, Phys. Rev. E {\bf67}, 021503 (2003).
\bibitem{ajay} A. D. Kachhvah, and N. Gupte, Phys. Rev. E {\bf83}, 036107 (2011).
\bibitem{copper} S. N. {Coppersmith}, C.-h. Liu, S. Majumdar, O. Narayan, and T. A. Witten, Phys. Rev. E {\bf 53}, 4673 (1996).
\bibitem{river} A. E. Scheidegger, Bull. Int. Acco. Sci. Hydrol. {\bf12}, 15 (1967).
\bibitem{griffeath} D. Griffeath, Additive and Cancellative Interacting Particles Systems, Vol. {\bf 724} of Lectures Notes in Mathematics (Springer-Verlag, Berlin, 1979); T. M. Liggett, Interacting Particles Systems (Springer-Verlag, Berlin, 1985).
\bibitem{domany} E. Domany and W. Kinzel, Phys. Rev. Lett. {\bf 53}, 311 (1984).
\bibitem{suki} B. Suki, A-L. Barab\'{a}si, Z. Hantos, F. Petak, and H.E. Stanley, Nature {\bf 368},615 (1994).
\bibitem{suki2} B. Suki, J. S. Andrade, M. F. Coughlin, D. Stamenovic, H.E. Stanley, M. Sujeer, and S. Zapperi, Annals of Biomedical Engineering {\bf 26}, 608 (1998).
\bibitem{achliop} D. Achlioptas, R. M. D'Souza, and J. Spencer, Science {\bf323}, 1453 (2009).
\bibitem{friedman} E. J. Friedman and A. S. Landsberg, Phys. Rev. Lett. {\bf103}, 255701 (2009).
\bibitem{radicchi} F. Radicchi and S. Fortunato, Phys. Rev. Lett. {\bf103}, 168701 (2009).
\bibitem{ziff} R. M. Ziff, Phys. Rev. Lett. 103, 045701 (2009); Phys. Rev. E {\bf82}, 051105 (2010).
\bibitem{araujo} N. A. M. Ara\'{u}jo and H. J. Herrmann, Phys. Rev. Lett. {\bf105}, 035701 (2010).
\bibitem{chen} W. Chen, and R. M. D'Souza,  Phys. Rev. Lett. {\bf106}, 115701 (2011).
\bibitem{moreira} A. A. Moreira, E. A. Oliveira,	S. D. S. Reis, H. J. Herrmann, and J.S. Andrade, Jr., Phys. Rev. E {\bf81}, 040101 (R) (2010).
\bibitem{chao} Y. S. Cho, B. Kahng, and D. Kim, Phys. Rev. E {\bf81}, 030103 (R) (2010).
\bibitem{dsouza} R. M. D'Souza and M. Mitzenmacher, Phys. Rev. Lett. {\bf104}, 195702 (2010).
\bibitem{daCosta} R. A. da Costa, S. N. Dorogovtsev, A. V. Goltsev and J. F. F. Mendes, Phys. Rev. Lett. {\bf105}, 255701 (2010).
\bibitem{riordan} O. Riordan and L. Warnke, Science {\bf333}, 322 (2011).
\bibitem{aval} A. D. Kachhvah and N. Gupte, Pramana - J. of Phys. {\bf 77}, No. 5, pp. 1-7 (2011).
\bibitem{riverdelta} H. Seybold, J. S. Andrade, Jr., and H. J. Herrmann, PNAS {\bf 104}, 16804 (2007).
\bibitem{marsgully} D. Reiss, G. Erkeling, K. E. Bauch, and H. Hiesinger, Geophysical Research Letters {\bf 37}, L06203 (2010).
\bibitem{shinbrot} T. Shinbrot, N.-H. Duong, L. Kwan, and M. M. Alvarez, PNAS {\bf 101}, 8542 (2004).
\bibitem{assym} In the natural structures, the channels of high capacity can have some channels 
of lower capacity joining them, leading to an overall symmetry between left and right connections, 
however one direction is favoured by the channels of high capacity, due to some feature like the nature 
of the geographic terrain, leading to variants of the $V-$ structure for the high capacity channels alone. 
\bibitem{foot} The above expression for capacity $w_i$ of a node $i$ is analogous to the expression for  fitness $\eta_i$ of a node $i$ in Bianconi $et$ $al.$ \cite{bianconi}.
\bibitem{bianconi} G. Bianconi and A.-L. Barab\'{a}si, Phys. Rev. Lett. {\bf86}, 05632 (2001).
\bibitem{huang} K. Huang, Statistical Mechanics (Wiley, Singapore, 1987).
\bibitem{footn} The slopes of the plots of $\big<n_w \big>$ against $w$ from simulation, for the original, the strategy-I, the strategy-II, and the V- lattices are found to be $0.9747$, $0.9954$, $0.9595$ and $0.9179$ respectively. The approximated slope values from Eq. $18$ for the original, the strategy-I, the strategy-II are $1.19$, $1.192$, and $1.22$ respectively.
\end{thebibliography}
\end{document}